\documentclass[twocolumn,showpacs,preprintnumbers,amsmath,amssymb,a4paper,superscriptaddress]{revtex4}
\usepackage{amsmath}
\usepackage{graphicx, subfigure}
\usepackage{dcolumn}
\usepackage{bm}

\begin{document}

\title{Shifts and widths of p-wave confinement induced resonances in atomic waveguides}
\date{\today}
\author{Shahpoor Saeidian}
\email[]{saeidian@iasbs.ac.ir}
\affiliation{Optics and Photonics Research Center, Department of Physics, Institute for Advanced Studies in Basic Sciences
(IASBS),
Gava Zang,
Zanjan 45137-66731,
Iran
}%
\author{Vladimir S. Melezhik}
\email[]{melezhik@theor.jinr.ru} \affiliation{Bogoliubov
Laboratory of Theoretical Physics, Joint Institute for Nuclear
Research, Dubna, Moscow Region 141980, Russian Federation}%
\affiliation{Department of Theoretical Physics, Dubna
International University for Nature, Society and Man, Dubna,
Moscow Region 141980, Russian Federation}%
\author{Peter Schmelcher}
\email[]{peter.schmelcher@physnet.uni-hamburg.de}
\affiliation{Zentrum f\"ur Optische Quantentechnologien,
Universit\"at Hamburg, Luruper Chaussee 149, 22761 Hamburg, Germany}%
\affiliation{The Hamburg Center for Ultrafast Imaging, Luruper Chaussee 149, 22761, Hamburg, Germany}%

\date{\today}
\begin{abstract}\label{txt:abstract}
We develop and analyze a theoretical model to study p-wave
Feshbach resonances of identical fermions in atomic waveguides by
extending the two-channel model of A.D. Lange et. al. [Phys. Rev. A
79, 013622 (2009)] and S. Saeidian et. al. [Phys. Rev. A 86, 062713
(2012)]. The experimentally known parameters of Feshbach
resonances in free space are used as input of the model. We
calculate the shifts and widths of p-wave magnetic Feshbach
resonance of $^{40}$K atoms emerging in harmonic waveguides as
p-wave confinement induced resonance (CIR). Particularly, we show
a possibility to control the width and shift of the p-wave
confinement induced resonance by the trap frequency and the applied magnetic field
which could be used in corresponding experiments. Our analysis
also demonstrates the importance of the inclusion of the effective radius in
the computational schemes for the description of the p-wave CIRs
contrary to the case of s-wave CIRs where the influence of this term
is negligible.
\end{abstract}

\maketitle

\section{INTRODUCTION}
The recent progress in ultracold atomic physics provides exceptional
possibilities for studying low-dimensional quantum systems. The
confining geometry of atomic traps can drastically alter their
ultracold scattering behaviour near the so-called confinement induced
resonances (CIRs) \cite{Olshanii1}.  Employing magnetic Feshbach
resonances \cite{Chin2010}, one can control the interaction between the atoms and
provide the conditions to experimentally observe the CIRs for
identical bosons \cite{Kinoshita, Paredes,Haller2009,Haller2010}
and fermions\cite{Guenter,Moritz,Kohl2011}, as well as
distinguishable atoms \cite{Lamporesi2010}.  CIRs have been
extensively studied, e.g. in the context of bosonic s-
\cite{Olshanii2, Melezhik07,Saeidian} and d-wave \cite{Giannakeas}
and dipolar \cite{Santos,Giannakeas13} scattering, fermionic p-wave scattering \cite{Granger,Melezhik07},
distinguishable atom scattering \cite{Kim2, Kim3, Melezhik07,
Melezhik09} in single-channel  as well as in multichannel (transversal excitation) regimes
\cite{Saeidian, Melezhik11,Olshanii3}. Beyond this
coupled l-wave CIRs in cylindrically symmetric waveguides have been explored \cite{Kim,Giannakeas12}.

In spite of the impressive progress concerning the experimental
\cite{Lamporesi2010,Haller2010,Kohl2011,Haller2011} and
theoretical \cite{Peng2010,Zhang2,Melezhik11,Sala2011}
investigations of CIRs the existing
theoretical models need to be improved for a quantitative
interpretation and guiding of the experiments. In the
seminal work of Olshanii \cite{Olshanii1} and in subsequent works
\cite{Olshanii2,LowDSystems,Yuro,Peng2010,Zhang2}, the simple form of a pseudo-potential was
used to model the interatomic interactions as compared to CIR
investigations with more realistic interatomic potentials
\cite{Kim2,Kim3,Saeidian,Melezhik09,Giannakeas,Melezhik11,Olshanii2,Granger,Sala2011,Grishkevich}.  However, all these
approaches are of single-channel (internal) character, allowing one to
explore only the main attribute of the Feshbach resonances in the
3D free space, i.e.  the appearance of a singularity in the s-wave
scattering length $a_s \rightarrow\pm\infty$ when a molecular
bound state with energy $E_B$ crosses the atom-atom scattering
threshold at energy $E=0$ in the entrance channel. However, other
important parameters of the Feshbach resonance, such as the
rotational and spin structure of the molecular bound state in the
closed channel as well as the width $\Delta$ of the resonance
characterizing the coupling $\Gamma$ of the molecular state with
the entrance channel, were ignored. In our  recent work
\cite{Saeidian2012} the single-channel scalar interatomic
interaction was replaced by the four-channel tensorial potential
modeling resonances of broad, narrow, and overlapping character
for bosons, according to the two-channel parametrization suggested
in \cite{Lange}. This allowed us to calculate the
shifts and widths of s-, d-, and g-wave magnetic Feshbach
resonances of Cs atoms emerging in harmonic waveguides as CIRs.

In the present work our particular focus are p-wave
collisions of identical fermions due to their anisotropic character.
Confinement induced shifts and widths of the resonances in an atomic
waveguide could be envisaged, similar to what has been studied for
s-wave interactions \cite{Saeidian2012}.  The main goal of the
present work is therefore to extend our model \cite{Saeidian2012} to the
fermionic case.  We then calculate the shifts and widths of the
Feshbach resonance for $^{40}$K atoms in its hyperfine state $|F=9/2,
m_F=-7/2\rangle$ for experimental conditions \cite{Guenter}.  The
parameters which can be obtained from the experiments on
magnetic Feshbach resonances in free space \cite{Ticknor,
Idziaszek}, namely, the spin characteristics, the background
scattering volume $V_{bg}$, the resonant energies $E_c$ (or the
corresponding resonant value of the strength $B_c$ of the external
magnetic field), and the width of the resonance $\Delta$
($\Gamma$) are used as input parameters of our model.

\section{FESHBACH RESONANCE MODEL IN FREE SPACE}
Let us first discuss an extension of our model
\cite{Saeidian2012}, developed for resonant s-wave scattering, to
the case of p-wave Feshbach resonances in 3D free space which
can be observed in a single-component Fermi
gas. Due to the Pauli exclusion principle the two-body wave
function must be antisymmetric, and therefore only odd partial
waves $l$ contribute to the fermionic scattering process in case of
a symmetric spin configuration.  In
contrast to the s-wave Feshbach resonance, the atoms near a p-wave
Feshbach resonance have to tunnel through a centrifugal barrier to
couple to the bound molecular state.  Therefore,
the wave function of the continuum can be significantly influenced by the bound state in the closed channel
only in a narrow range of the magnetic field.
Another feature of the p-wave Feshbach resonances is a doublet
structure which arises from the magnetic dipole-dipole interaction
between the electronic spins of the atoms.  This splits the
Feshbach resonance into distinct resonances indicated by their
partial-wave projection $m_l=0$ and $|m_l|=1$ along the
quantization axis.

In this work we investigate collisions of two spin-polarized
$^{40}$K atoms in the hyperfine state $|F=9/2, m_F=-7/2\rangle$.
One can write the joint state of the atom pair as $|f_1=9/2,
m_{f_1}=-7/2\rangle|f_2=9/2, m_{f_2}=-7/2\rangle |l=1,
m_{l}\rangle$ where $m_l=0, \pm1$. The dipole-dipole interaction
between the valence electrons of the atom pairs can be written as
\begin{equation}\nonumber
H_{ss}=-\alpha^2\frac{3(\hat{\textbf{r}}\cdot\hat{S}_1)(\hat{\textbf{r}}\cdot\hat{S}_2)-(\hat{S}_1\cdot\hat{S}_2)}{r^3}
\end{equation}
where $\alpha$ is the fine structure constant, $\hat{S}_i$ is the
spin of the valence electron of $i$th atom, $r$ is the interatomic
separation, and $\hat{\textbf{r}}$ is the unit vector defining the
direction of interatomic axis.  It couples different partial wave
components $|l, m_l\rangle$ and $|l', m_l'\rangle$ with $l'=l\pm2$
or $l'=l\neq0$ \cite{Ticknor}, and breaks the rotational
invariance.  The different components $m_l$ contribute differently
in the dipole-dipole interaction, which means that molecular bound
states with different $m_l$ possess different energies \cite{Ticknor}.
Consequently, the Feshbach resonances for different $m_l$ values
couple to distinct molecular bound states and thus have different
magnetic-field dependence.

The difference in the pair collisions with the different $m_l$
projections can be understood intuitively by considering the
dipole-dipole interaction between the two atoms. The external
magnetic field orients the spin of the valence
electron of $^{40}$K in $|F=9/2, m_F=-7/2\rangle$ state.  When two
dipoles are aligned side-by-side (head-to-tail), they are in a
repulsive (attractive) configuration, corresponding to
$(\hat{\textbf{r}}\cdot\hat{S_i})=0$
($(\hat{\textbf{r}}\cdot\hat{S_i})=1$).

The difference between the states $m_l=0$ and $|m_l|=1$ is
illustrated in Fig.3 of ref.\cite{Ticknor}. In the case the motion
is in a plane containing the magnetic field the  interaction alternates between attractive and
repulsive as the dipoles change between head to tail attraction
and side by side repulsion.  On the other hand, for the case that
the atoms move in the plane perpendicular to the magnetic field
the dipoles are held in the side-by-side configuration, and the interaction is only repulsive.
Due to this effect one expects that the dipole-dipole interaction
lifts the degeneracy between the $m_l=0$ and $|m_l|=1$ collisional
channels which leads to a splitting of the Feshbach resonances, with
corresponding resonance for $|m_l|=1$ at a higher energy
\cite{Ticknor}.  This doublet structure disappears with increasing
temperature due to a broadening of the resonance.

For p-wave collisions the relevant parameter is the scattering volume, $V_p$
which has been parameterized as \cite{Ticknor}

\begin{equation}\label{scattering volume}
k^3 cot\delta_1(k, B)=-\frac{1}{V_p(B)} + \frac{k^2}{R(B)}
\end{equation}
near the magnetic Feshbach resonance, where $V_p$ and the effective range $R$ were approximated as a quadratic function of the magnetic field B
\begin{equation}
\frac{1}{V_p(B)} = c^{(0)} + c^{(1)}B + c^{(2)}B^2
\end{equation}
\begin{equation}
\frac{1}{R(B)} = d^{(0)} + d^{(1)}B + d^{(2)}B^2 \,\,.
\end{equation}
The coefficients $c^{(i)}$ and $d^{(i)}$ were obtained by fitting
to the experimental data and are given in Table I of \cite{Idziaszek}.
A simple formula for an energy-dependent scattering volume has been
derived in the framework of multichannel quantum-defect theory
(MQDT)
\begin{equation}\label{scattering volume of E}
V_p(E,B) = V_{bg}(E)\left[1-\frac{\Delta(1+\frac{E^3}{E_{bg}^3})}{B-B_0-\frac{E}{\delta\mu}+\frac{E^3}{E_{bg}^3}} \right] \,\,,
\end{equation}
where $E_{bg}=\frac{\hbar^2}{2\mu a_{bg}^2}$ is the energy associated with the background scattering length $a_{bg}=\lim_{E\rightarrow 0}[V_{bg}(E)]^{1/3}$ ($\mu =m/2$ is the reduced mass).  Except for the case $a_{bg}\gg R_{vdW}$, $E_{bg}\sim E_{vdW}$, and therefore $E\ll E_{vdW}$ (here $R_{vdW}=1/2(2\mu C_6/\hbar^2)^{1/4}$ is the van der Waals radius,
 $C_6$ is the corresponding van der Waals coefficient and $E_{vdW}=\hbar^2/2\mu R_{vdW}^2$ ), when one does not need to take into account the effective range expansion of $V_{bg}(E)$, Eq.(\ref{scattering volume of E}) can be simplified to
\begin{equation}\label{scattering volume of E simplified}
V_p(E,B) = V_{bg}\left[1-\frac{\Delta}{B-B_0-\frac{E}{\delta\mu}}\right]
\end{equation}
Fitting the expansion  (\ref{scattering volume}) to this formula, one obtains the resonance parameters listed in Table I \cite{Idziaszek}.

\begin{table}
\begin{tabular}{|c|c|c|c|c|c|c|}
\hline
    &   &   &   &   \\
  $|m_l|$ & $B_0 [G]$ & $B^* [G]$ & $\frac{\Gamma}{h} [MHz]$ & $\Delta [G]$ & $\frac{\delta\mu}{h} [\frac{kHz}{G}]$ & $a_{bg} [a_0]$ \\
   &   &   &   &   \\
  \hline
  0 & 198.85 & 178.508 & 0.5684 & -20.342 & 93.093 & -104.26 \\
  \hline
  1 & 198.37 & 175.414 & 0.6950 & -22.956 & 92.667 & -99.539 \\
  \hline
\end{tabular}
\caption{The poles $B_0$ and zeros $B^*$ of the scattering valume
$V_p(B)$, the coupling strength $\Gamma$, the resonance width
$\Delta$, the magnetic moment difference $\delta\mu$ and the
background scattering length $a_{bg}$ for a p-wave Feshbach
resonance between two $^{40}$K atoms in the hyperfine state $|F=9/2,
m_F=-7/2\rangle$ for two relative angular momentum
projections $m_l$ along the axis of the magnetic field.} \label{tabI}
\end{table}

Using MQDT in the two channel case, the following formula for the phase shift in the open channel
\begin{equation}\label{phase shift}
\delta(E,l) = \delta_{bg}(E,l) - \arctan\left(\frac{\frac{\Gamma}{2}C^{-2}(E,l)}{E-E_c+\frac{\Gamma}{2}\tan\lambda(E,l)}\right)
\end{equation}
was obtained in ref.\cite{Idziaszek}. The first term $\delta_{bg}(E,l)$ is the phase shift resulting from the scattering in the open channel only,
i.e. the background phase shift.  The second term describes the resonant contribution originating from a bound state in the closed channel with the energy $E_c$ located close to the threshold of the open channel.  $C^{-2}(E)$ and $\tan\lambda(E)$ are MQDT functions.  The width of the resonance $\Gamma$ is multiplied by $C^{-2}(E)$ which accounts for a proper threshold behaviour as $k\rightarrow 0$.

We assume that the bound state can be linearly tuned by a magnetic Zeeman shift, i.e. $E_c(B)=\delta\mu(B-B_c)$, where $\delta\mu$ is the difference of
the magnetic moments of the open and closed channels, and $B_c$ is the crossing field value of the bound state.  In the case of a power law interaction potential $r^{-s}$ the phase shift and MQDT function exhibit the following Wigner threshold behavior as $E\rightarrow  0$ \cite{Idziaszek}
\begin{equation}\label{threshold1}
\delta_{bg}(E) \longrightarrow A_lk^{2l+1}
\end{equation}
for $   2l+1\leq s-2$  and
\begin{equation}\label{threshold2}
C^{-2}(E) \longrightarrow B_lk^{2l+1}
\end{equation}
and
\begin{equation}\label{threshold3}
\tan \lambda(E) \longrightarrow \tan \lambda(0)
\end{equation}
for all $l$.  For van der Waals forces the higher order terms in $k$ can be neglected in case $E\lesssim E_{vdW}$.
For alkali atoms $E_{vdW}$ ranges from 0.1mK to 30mK \cite{Julienne}. Hence, in the ultracold regime ($E\lesssim 1\mu K$), one can safely use the approximation (\ref{threshold1})-(\ref{threshold3}).  The width $\Delta$ of the magnetic Feshbach resonance reads
\begin{equation}\label{width}
\lim_{E\rightarrow 0} \frac{\Gamma}{2}\frac{C^{-2}(E)}{\tan\delta_{bg}(E)}=-\delta\mu\Delta
\end{equation}
and the resonance position $B_0$, that is shifted from $B_c$ due to the coupling between the open and closed channels
take on the following appearance
\begin{equation}\label{B-zero}
B_0=B_c + \frac{\Gamma}{2\delta\mu}\lim_{E\rightarrow 0}\tan \lambda(E).
\end{equation}

 Further, using the above parametrization, we extend the scheme suggested in ref.\cite{Lange} to describe the magnetic Feshbach resonances of the p-wave scattering in an ultracold $^{40}$K gas.
The two-body problem in free space permits the separation of the
center-of-mass and relative motion yielding the following
Hamiltonian for the relative atomic motion
\begin{equation}\label{H3D}
\hat{H}(r,\theta)=\left[-\frac{\hbar ^2}{2\mu}\nabla^2 \right]\hat{I} + \hat{V}(r)
\end{equation}
Here $\hat{V}(r)$ is the two-channel interatomic potential and
$\hat{I}$ is the unit matrix and $r$ is the relative radial coordinate. Let
us suppose that initially the spin-polarized atoms are prepared in
the entrance channel $|e\rangle$ and the ``closed channels"
$|c\rangle$ supports molecular bound states at $B_0$. The quantum
state of an atomic pair with energy $E$ is then described as
$|\psi\rangle=\psi_{c}({\bf r})|c\rangle+\psi_{e}({\bf
r})|e\rangle$ satisfying the Schr\"{o}dinger equation with the
Hamiltonian (\ref{H3D}).  A two-channel square-well potential
\begin{equation}\label{Tensor potential}
\hat{V}=\left (\begin{array}{cc} -V_{c}& \hbar\Omega\\ \hbar\Omega & -V_e
\end{array}\right )  \begin{array}{cc} (if & r<\overline{a})\end{array}\\
\end{equation}
\begin{equation}\nonumber
=\left (\begin{array}{cccc} \infty& 0 \\ 0 & 0
  \end{array}\right )  \begin{array}{cc} (if & r>\overline{a})\end{array}
\end{equation}
is employed to describe the colliding atoms in the ``entrance
channel" $|e\rangle$ and the weakly-bound molecule in the ``closed
channel" $|c\rangle$ near a Feshbach resonance.  The coupling
constant $\hbar\Omega$ induces Feshbach couplings between the
channels. For $r< \overline{a}$, we assume that the attractive
potential supports multiple molecular states - that is $
V_{e}$,$ V_{c}\gg E_{vdW}=\hbar^2/2\mu R_{vdW}^2$
($R_{vdW}=65a_0$ and $a_0$ is the Bohr
radius). For $r> \overline{a}$, entrance- and closed-channel thresholds
are set to be $ E=0$ and $E=\infty$, respectively. Here $\bar{a}=(\bar{V})^{1/3}$ and
$\bar{V}=\frac{1}{3\sqrt{2}}\frac{\Gamma(1/4)}{\Gamma(7/4)}R_{vdW}^3$
is the mean scattering volume \cite{Gautam}, and $\Gamma(x)$ is
the gamma function.  For the pair of $^{40}$K atoms we have the value $\bar{a}=63.4a_0$.

Such a choice of the interatomic interaction permits a simple
parametrization of the atom-atom scattering in universal terms of
the energy of the bare bound state $E_c$,  the Feshbach coupling
strength $\Gamma$ of the bound molecular state with the entrance channel
and the background scattering volume $V_{bg}$, which is
convenient for an analysis of experimental data near magnetic
Feshbach resonances \cite{Chin2010,Lange}.

When the mixing between the closed channel and the entrance
channel is weak and the background scattering length $|a_{bg}|$
considerably exceeds the range of the interatomic interaction
$\bar{a}$, the p-wave scattering volume $V_p$ obeys the following
expression
\begin{equation}
\frac{1}{V_p-\bar{V}}=\frac{1}{V_{bg}-\bar{V}}+\frac{\Gamma/2}{\bar{V}E_c}
\end{equation}
The parameters $V_{bg}$, $\delta\mu$, $B_0$, and $\Delta$ from
\cite{Idziaszek} together with $V_p(B)$ given by
Eq.(\ref{scattering volume of E simplified}) are used for fitting
the diagonal terms $V_c$ and $V_e$ in the tensor potential
(\ref{Tensor potential}).  The nondiagonal terms $\hbar\Omega$ are
defined by $\Gamma/2=2\theta^2V_c$, where $\tan
2\theta=2\hbar\Omega/(V_e-V_c)$.

The scattering volume $V_p(B)$ is then calculated for different
$B$ and varying parameters of the potential $\hat{V}$ by solving
the Schr\"{o}dinger equation
\begin{equation}\label{SCHEq}
\left( \left[-\frac{\hbar^2}{2\mu}\nabla^2 \right] \hat{I}+\hat{B}+\hat{V}(r) \right) |\psi\rangle=E|\psi\rangle
\end{equation}
with scattering boundary condition
\begin{equation}
\psi_e(r)\rightarrow e^{ikz}+\frac{f(k, \theta)}{r}e^{ikr},\psi_c(r)\rightarrow 0
\end{equation}
at $kr\rightarrow\infty$ for the fixed $E\rightarrow 0$ ($k=\sqrt{2\mu E}/\hbar$) \cite{MelHu}.
The diagonal matrix $\hat{B}$ in Eq.(\ref{SCHEq}) is defined as  $B_{cc}=\delta\mu(B-B_c)$ and $B_{ee}=0$.  After separation of the angular part in Eq.(\ref{SCHEq}) we come to the system of two coupled radial equations
\begin{equation}
\left[-\frac{\hbar^2}{2\mu}\frac{d^2}{dr^2} +\frac{\hbar^2l(l+1)}{2\mu r^2} +B_{ii}
\right] \phi_i(r) + \sum_{j}V_{ij}(r)\phi_{j}(r)= E\phi_{i}(r)
\end{equation}
with $l=1$, and $i, j=e, c$.

By setting $\frac{C^{-2}(E)}{\tan\delta_{bg}(E)}
=-\frac{V_{bg}\bar{V}}{(V_{bg}-\bar{V})^2}$ and varying $V_c$,
$V_e$ and $\Omega$ we obtain an excellent agreement of the
calculated p-wave energy dependent scattering volume
$V_p(B,E)=-k^{-3}\tan\delta_1(B,k)$ (Eq.(\ref{scattering volume}))
with the expression (\ref{scattering volume of E simplified}) from
\cite{Idziaszek} for $^{40}$K atom in the hyperfine state $|F=9/2,
m_F=-7/2\rangle$ for $196G<B<202G$. The found zeros $B^*$ and
poles $B_0$ of the scattering volume and the coupling strength
$\Gamma$ are given in Table I. Here $B_c=B_0+\beta\Delta$,
$\Delta=B^*-B_0$, and $\beta=V_{bg}/V_{bg}-\bar{V}$. We 
note that such a procedure yields a $k^3\cot\delta_1(B,k)$ coinciding
with the experimental data \cite{Ticknor} for low colliding energies
at $T\sim 1-10 nK$. The agreement, however, weakens with increasing $k$. Therefore, our
$1/R=-1/2\frac{d^2}{dk^2}(1/V_p(B,E))$ differs from the value
given in \cite{Ticknor} by about a factor of 3. This fact is,
however, understandable because we did not use $1/R$ directly
from \cite{Ticknor} in our fitting procedure for obtaining the
parameters of the interaction potential (\ref{Tensor potential}).

\begin{figure}
\includegraphics[width=1.0\columnwidth]{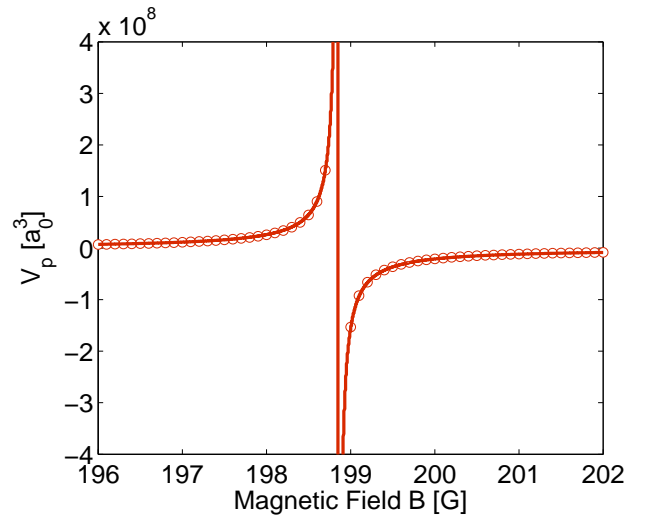}
\caption{(Color online) The p-wave scattering volume $V_p$
calculated for potassium atoms near the Feshbach resonance in
the hyperfine state $|F=9/2, m_F=-7/2\rangle$ and the relative angular momentum state $|l=1, m_l=0\rangle$ as a function of the magnetic field $B$. The solid curve shows the analytical result (5) and the dots show the numerical result (see text).}
\label{fig1}
\end{figure}

Fig.\ref{fig1} shows our results obtained for the state $m_l=0$ for
the scattering volume as a function of the magnetic field $B$ at
the temperature $T=1.0 nK$. For comparison we have also plotted
the analytical curve (5) which is in a good agreement with the
numerical result. The computations were performed for
$V_e=1.5\times 10^{-2} [a.u.]$, $V_c=5.9\times 10^{-2} [a.u.]$ and
$\Omega=1.2\times 10^{-6} [a.u.]$ modeling unbound two-atom states
in the entrance channel $|e>$ and the resonance state with the
parameters fixed for $m_l=0$ in the Table I.

\section{FESHBACH RESONANCE MODEL IN A WAVEGUIDE}

Next we analyze the scattering properties of the p-wave magnetic
Feshbach resonances in harmonic waveguides permitting unbound
motion in the longitudinal $z$-direction and a transversally strongly
confined $\rho$-motion in the potential
$1/2\mu\omega_{\perp}^2\rho^2$. To describe
the scattering process in the waveguide we have to calculate the
scattering amplitude $f_p(E)$ by integrating the Schr\"odinger
equation
\begin{equation}\label{SCHEqin waveguides}
\left( \left[-\frac{\hbar^2}{2\mu}\nabla^2+\frac{1}{2}\mu\omega_{\perp}^2\rho^2 \right]
\hat{I}+\hat{B}+\hat{V}(r)\right)|\psi\rangle=E|\psi\rangle
\end{equation}
with the scattering boundary conditions
\begin{equation}\label{boundary condition}
\psi_e({\bf r}) = \left (\sin (k_0 z) + sgn(z)f_p e^{ik_0 \mid
z\mid}\right )\Phi_{0}(\rho)\,\,,\,\,\, \psi_{c,i}({\bf
r})\rightarrow 0
\end{equation}
at $\mid z\mid = \mid r\cos\theta\mid \rightarrow \infty$ adopted
for a confining trap \cite{Saeidian}. Here, $\Phi_0(\rho)$ is the
ground-state wave-function of the two-dimensional harmonic
oscillator  and $k_0=\sqrt{2\mu
(E-\hbar\omega_{\perp})}/\hbar=\sqrt{2\mu E_{\parallel}}/\hbar$.
We consider pair collisions of identical fermionic potassium
atoms, therefore the scattering wave-function $|\psi\rangle$ must
be antisymmetric with respect to the exchange $z \rightarrow -z$.
In the presence of the harmonic trap
$1/2\mu\omega_{\perp}^2\rho^2$ the azimuthal angular part of the
solution is separated, and Eq.(\ref{SCHEqin waveguides}) is
reduced to the coupled system of two 2D Schr\"{o}dinger-type
equations in the plane \{$\rho$, $z$\}. To integrate  this coupled
channel scattering problem in the plane $\{r,\theta\}$ we have
extended the computational scheme developed in \cite{Saeidian}. The integration
was performed in the units of the problem leading to the scale
transformation: $r \rightarrow \frac{r}{\overline{a}}$, $E
\rightarrow \frac{E}{E_0}$, $V \rightarrow \frac{V}{E_0}$, and
$\omega_{\perp} \rightarrow \frac{\omega_{\perp}}{\omega_0}$ with
$E_0=\frac{\hbar^2}{\mu\overline{a}^2}$ ($=6.8\times 10^{-9} E_h$
for $^{40}$K atoms), and $\omega_0=\frac{E_0}{\hbar}$ ($=2.8\times
10^{5} kHz$ for $^{40}$K atoms).

\section{COMPUTATIONAL APPROACH}
First, we discretize the Schr\"{o}dinger Eq.(\ref{SCHEqin waveguides}) on a grid of the angular variable
$\{\theta_j\}^{N_{\theta}}_{j=1}$, which has been introduced in
\cite{Saeidian}.  We expand the solution of Eq.(\ref{SCHEqin waveguides}) in the basis
$g_j(\theta)=\sum^{N_{\theta}-1}_{l=0}P_l(\cos\theta)(\hat{P}^{-1})_{lj}$
according to
\begin{equation}\label{expanded wave}
|\psi\rangle =\sum_{i=e,c}\psi_i(r,
\theta)=\frac{1}{r}\sum_{i=e,c}\sum^{N_{\theta}}_{j=1}g_j(\theta)u^{(i)}_j(r)\,\,,
\end{equation}
where $\hat{P}^{-1}$ is the inverse of the $N_{\theta}\times
N_{\theta}$ matrix $\hat{P}$ with elements defined as
$P_{jl}=\lambda_jP_l(\cos\theta_j)$ and $\lambda_j$ are the
weights of the Gauss quadrature.  Substituting (\ref{expanded
wave}) into (\ref{SCHEqin waveguides}) results in a system of
$2 N_{\theta}$ Schr\"{o}dinger-like coupled equations with
respect to the $2\times N_{\theta}$ dimensional unknown vector
$\textbf{u}(r)=\{u_{ji}(r)\}=\{\lambda_j^{1/2}u^{(i)}_j(r)\}^{N_{\theta}}_{j=1,
i=e,c}$
\begin{equation}
[\hat{K}(x)+2(E\hat{I}-\hat{U}(r(x))-\hat{B})]\textbf{u}(r(x))=0
\end{equation}
with
$$
K^{ii'}_{jj'}(x)=[\delta_{jj'}\beta^2(x)(\frac{d^2}{dx^2}-\gamma\frac{d}{dx})-
$$
\begin{equation}
\frac{1}{r^2}\sum^{N_{\theta}-1}_{l=0}P_{jl}l(l+1)(P^{-1})_{lj'}]\delta_{ii'}
\end{equation}
and
\begin{equation}
U^{ii'}_{jj'}(r)=[V_{ii'}(r)+\frac{1}{2}\omega_{\perp}^2\rho_j^2\delta_{ii'}]\delta_{jj'}\,,\,
\rho_j=r\sin\theta_j\,.
\end{equation}
with $\beta(x)= (e^{\gamma}-1)/(r_m
\gamma e^{\gamma x})$ and we mapped and discretized the initial
variable $r\in(0,r_m]$ onto the uniform grid $x_j\in(0,1]$
according to
\begin{equation}
r_j=r_m\frac{e^{\gamma x_j}-1}{e^{\gamma}-1}, j=1, 2, ..., N \,\,,
\end{equation}
where $r_m$ is chosen in the asymptotic region $r_m\rightarrow
\infty$ and $\gamma > 0$ is a tuning parameter.  Using the finite
difference approximation, the above Schrodinger-like equations are
reduced to a system of algebraic equations
according to ref. \cite{Saeidian}. By solving this system of equations for
fixed colliding energy $E$ and matching the calculated vector
$\psi_e(E, r=r_m, \theta_j)$ with the asymptotic behavior
(\ref{boundary condition}) at $r=r_m$, we find the scattering
amplitude $f_p$, from which one can obtain the observable
transmission coefficient $T(E) = |1 + f_p(E)|^2$ in the waveguide.

\begin{figure}[ht!]
\begin{center}
\begin{subfigure}{
\includegraphics[width=0.7\columnwidth]{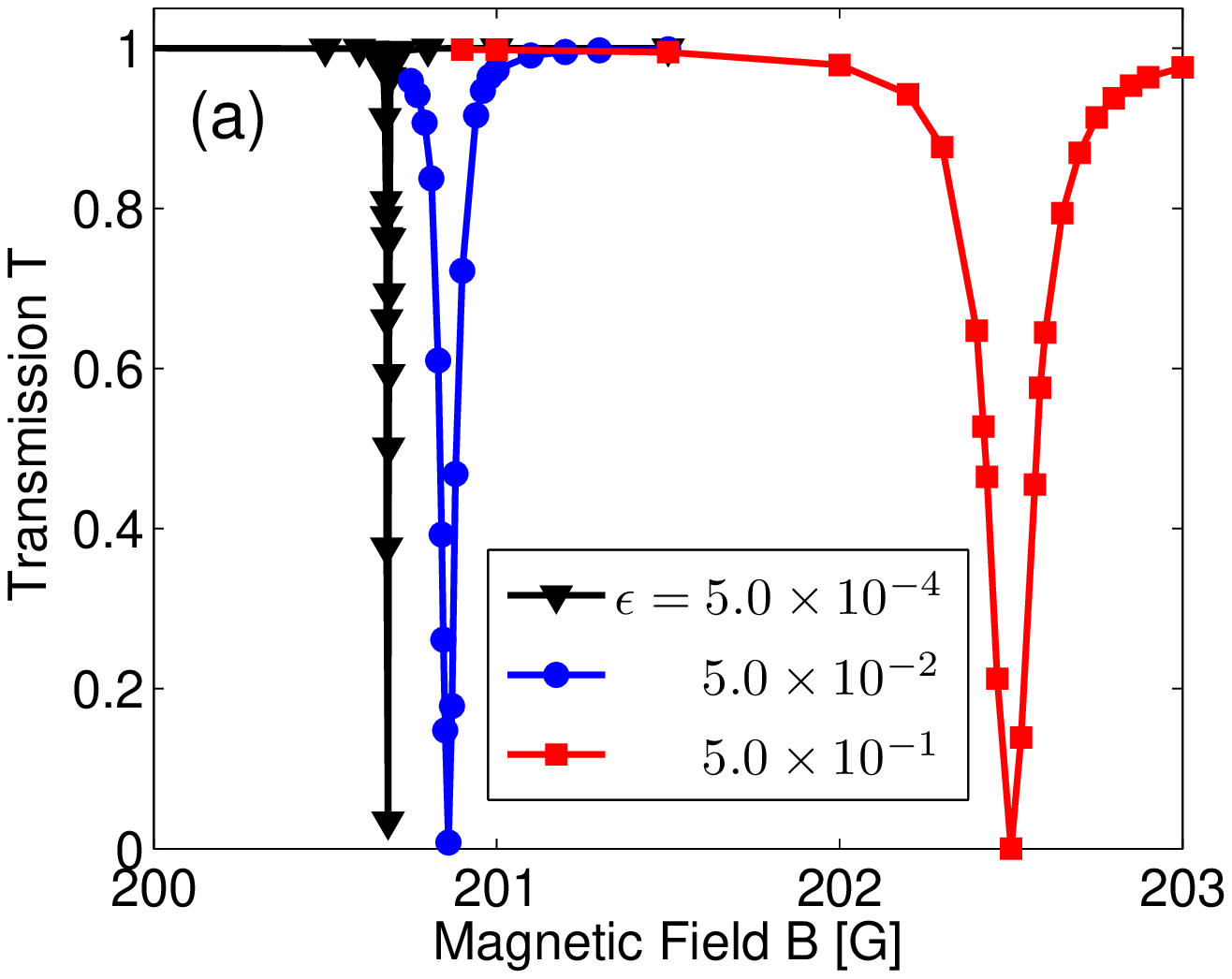}}
\end{subfigure}\\
\begin{subfigure}{
\includegraphics[width=0.7\columnwidth]{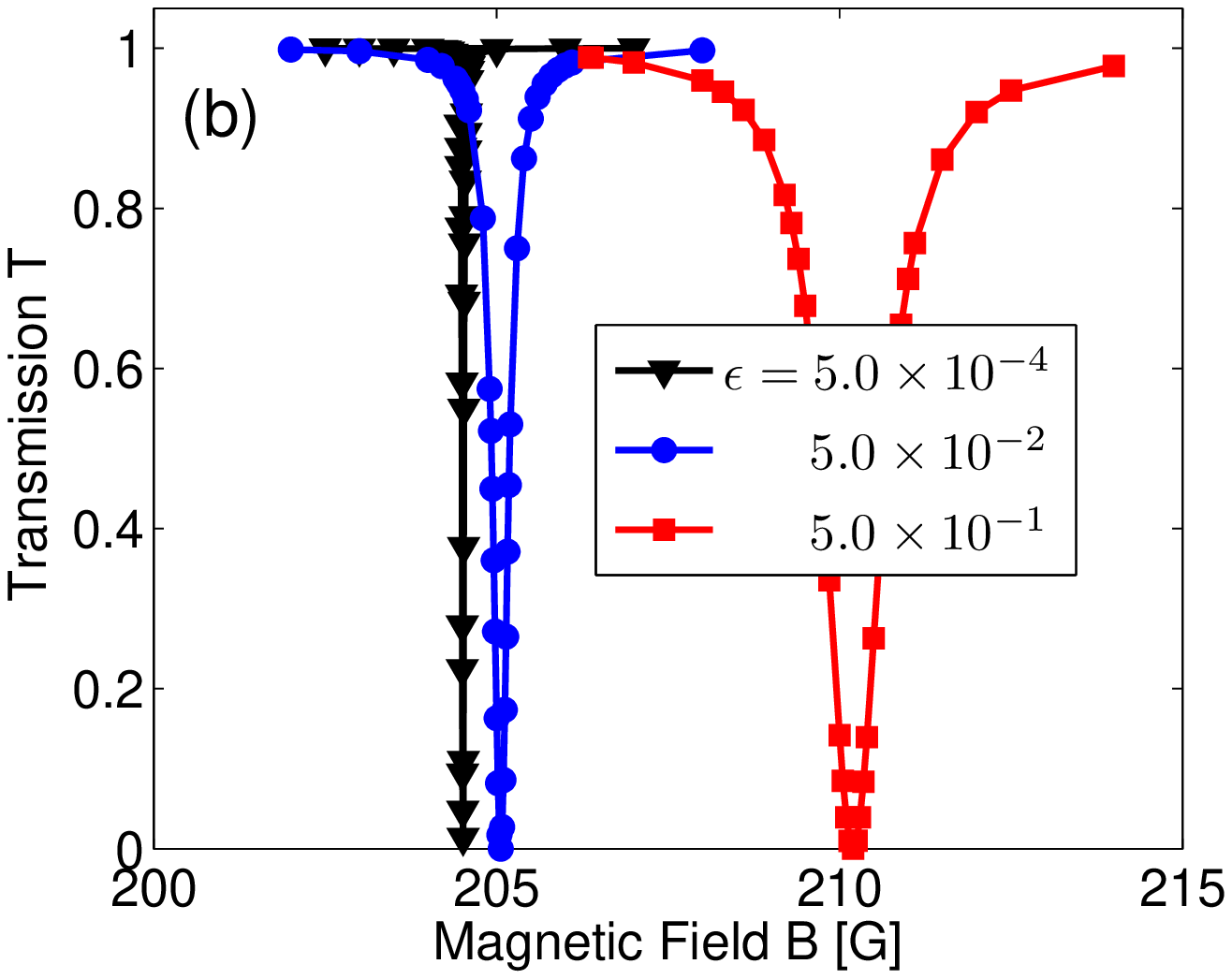}}
\end{subfigure}
\end{center}
\caption{(Color online) The transmission $T$ in the harmonic waveguide with (a) $\omega_{\perp}=0.002$ and (b)
$\omega_{\perp}=0.006$ as a function of external magnetic field $B$ for several rescaled longitudinal energies $\epsilon$.}
\label{fig2}
\end{figure}

\begin{figure}
\includegraphics[width=1.\columnwidth]{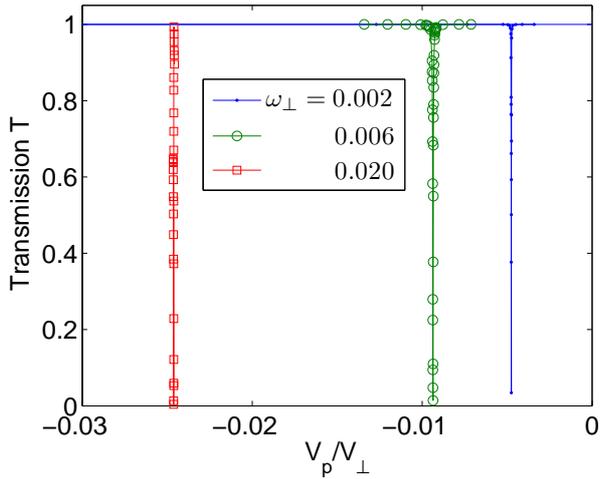}
\caption{(Color online) Transmission $T$ in the waveguide for the low energy limit $\epsilon=5\times 10^{-4}$ as a
function of $V_p/V_{\perp}$.} \label{fig3}
\end{figure}

\begin{figure}
\includegraphics[width=1.0\columnwidth]{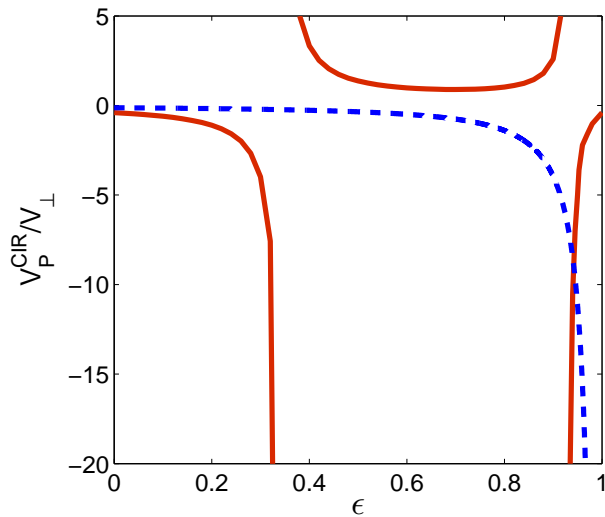}
\caption{(Color online) The position of the p-wave CIR in the
harmonic waveguide as a function of $\epsilon$ predicted by
Eqs.(\ref{resonance position old formula}) (solid curve) and
(\ref{resonance position Kim}) (dashed curve).}
\label{fig4}
\end{figure}

\begin{figure}
\includegraphics[width=1.\columnwidth]{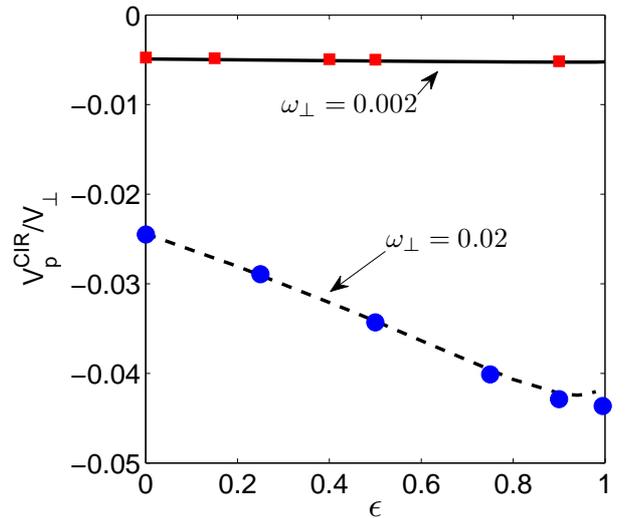}
\caption{(Color online) The position of the p-wave CIR as a function of $\epsilon$ in harmonic waveguides
with $\omega_{\perp} = 0.002$ and $0.02$. The curves show the
analytical results (\ref{resonance position new formula}) and the
dots correspond to the numerical results.} \label{fig5}
\end{figure}

\begin{figure}
\includegraphics[width=1.0\columnwidth]{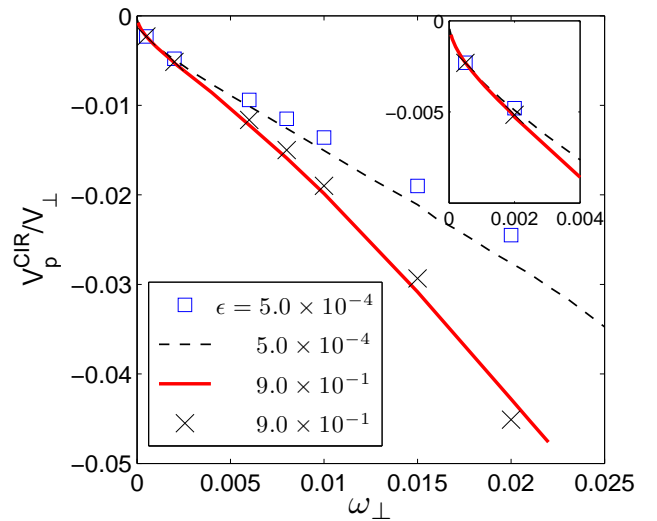}
\caption{(Color online) The position of the p-wave CIR in the harmonic waveguide as a function of
$\omega_{\perp}$.  The curves show the analytical results (\ref{resonance position new formula}) and the dots correspond to the
numerical results.} \label{fig6}
\end{figure}

\begin{figure}
\includegraphics[width=1.\columnwidth]{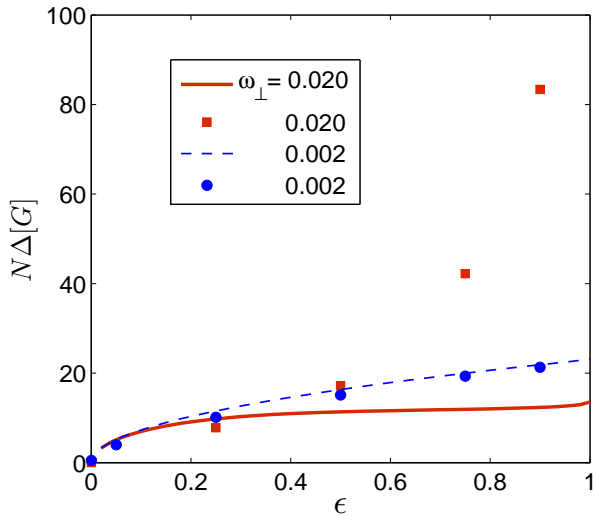}
\caption{(Color online) The width of the p-wave CIR of potassium
atoms in the hyperfine state $|F=9/2, m_F=-7/2\rangle$ and the relative angular momentum state $|l=1, m_l=0\rangle$ as a function of the rescaled energy $\epsilon$ in harmonic waveguides
with $\omega_{\perp} = 0.002$ and  $0.02$. $N=100$ and $1$
for $\omega_{\perp} = 0.002$ and  $0.02$ respectively. The curves
show the analytical results (\ref{width formula}) and
the dots show the numerical results.} \label{fig7}
\end{figure}

\begin{figure}
\includegraphics[width=1.\columnwidth]{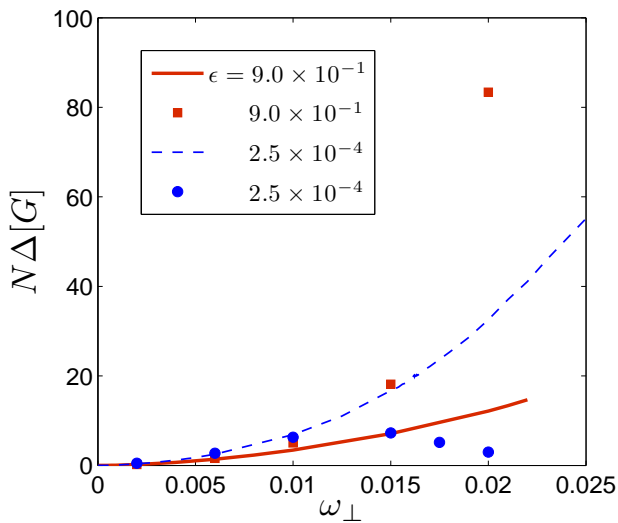}
\caption{((Color online) The width of the p-wave CIR of potassium
atoms confined by harmonic waveguide in the hyperfine state $|F=9/2, m_F=-7/2\rangle$ and the relative angular momentum state $|l=1, m_l=0\rangle$ as a function of the waveguide
frequency $\omega_{\perp}$ for $\epsilon = 5.0\times 10^{-4}$ and
$9.0\times 10^{-1}$.  $N=100$ and $1$ for $\epsilon =
5.0\times 10^{-4}$ and $9.0\times 10^{-1}$ respectively. The
curves show the analytical results (\ref{width formula}) and the dots show the numerical results.} \label{fig8}
\end{figure}

\begin{figure}
\includegraphics[width=1.\columnwidth]{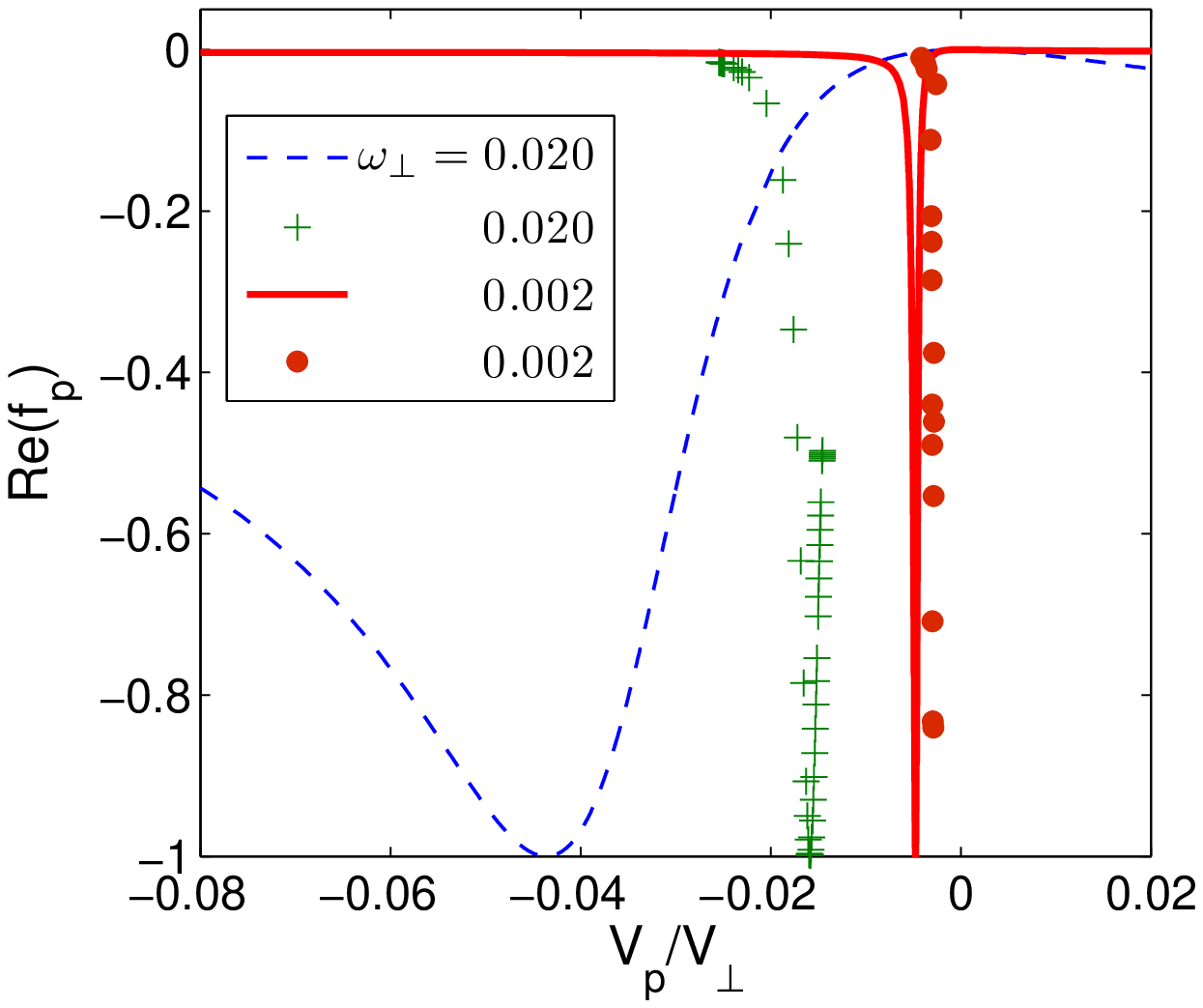}
\caption{(Color online) The real part of the scattering amplitude $f_p$ (dots) as a function of the $V_p/V_{\perp}$
along with the analytical results (\ref{scattering amplitude formula}) (curves) for $\epsilon=0.9$.}
\label{fig9}
\end{figure}

\section{RESULTS AND DISCUSSION}
\subsection{Transmission coefficient}
We have performed two sets of integration of Eq.(\ref{SCHEqin waveguides}) for varying interatomic interaction (defined by a
varying magnetic field $B$) of $^{40}K$ atoms
in the hyperfine state $|F=9/2, m_F=-7/2\rangle$ and the relative angular momentum state $|l=1, m_l=0\rangle$ near the Feshbach resonance $B_0=198.85G$: the
dependence of the transmission
$T(B,\omega_{\perp},\epsilon)=|1+f_p(B,\omega_{\perp},\epsilon)|^2$
on the trap frequency $\omega_{\perp}$, defining the transversal
volume $V_{\perp}=(\hbar/(\mu\omega_{\perp}))^{3/2}$, and the
rescaled longitudinal energy
$\epsilon=\frac{E_{\|}}{2\hbar\omega_{\perp}}=\frac{E-\hbar\omega_{\perp}}{2\hbar\omega_{\perp}}=\frac{E}{2\hbar\omega_{\perp}}-\frac{1}{2}$
was analyzed. Calculations were performed with the parameters
$V_e$, $V_c$, and $\Omega$ of the interaction potential $\hat{V}$
(\ref{Tensor potential}) fixed in Section II by fitting the
parameters of the Feshbach resonance of potassium atoms in free space.

In Fig.\ref{fig2} the dependence of the transmission
$T(B,\omega_{\perp},\epsilon)$ on the external magnetic field $B$
is given for two different $\omega_{\perp}$ and varying colliding
energy. The p-wave Feshbach resonance is manifested in the
waveguide as a minimum in the transmission $T(B_{CIR})=0$ (i.e. as
a p-wave CIR)\cite{Granger,Kim,Saeidian}. Fig.\ref{fig2}
demonstrates the strong dependence of the shift of the resonance
$B_0-B_{CIR}$ in the confining trap on the trap frequency
$\omega_{\perp}$ similar to the s-wave CIRs \cite{Saeidian2012}. This
effect, as in the case of the s-wave CIR \cite{Haller2009,Haller2010}, can be used to control the
p-wave CIR position. However, for a complete control of the p-wave
resonance shift in the waveguide one has also to understand the
strong dependence on the energy $\epsilon$ shown in
Fig.\ref{fig2}. This effect has to be contrasted to the position of
the s-wave CIR given by $a_s(B_{CIR})/a_{\perp}= 0.68$ (where $a_{\perp}= V_{\perp}^{1/3}$
and $a_s$ is the s-wave scattering length) \cite{Olshanii1} which is taken in the
$\epsilon \rightarrow 0$ limit and was confirmed in recent
experiments \cite{Haller2009,Haller2010}. The strong dependence of
the p-wave CIR position on $\epsilon$ is equally known \cite{Granger}. However, subsequent
investigations, both of analytical \cite{Kim3,Peng2014} and numerical character
\cite{Saeidian}, have demonstrated strong deviations from
the results in ref. \cite{Granger}. In the next subsection we shall discuss
the origin of this difference. To this end we study the transmission
$T(V_p(B),\omega_{\perp},\epsilon)=\mid 1+f_p(V_p(B),\omega_{\perp},\epsilon)\mid^2$ as a function of the
p-wave scattering volume $V_p(B)/V_{\perp}$ ``normalized'' to the
transversal volume $V_{\perp}$ and as a function of the trap fequency $\omega_{\perp}$
in the low energy limit, which is shown in Fig.\ref{fig3}.
The dependence of the p-wave CIR position $V_p^{CIR}/V_{\perp}$, defined as
$T(V_p^{CIR}/V_{\perp})=0$, on $\omega_{\perp}$ qualitatively
shows the same behaviour as in our previous work \cite{Saeidian} .
However, the calculated values of $|V_p^{CIR}/V_{\perp}|$ in \cite{Saeidian}
are approximately two times larger than the present values.
This difference stems from the definition of $V_p$ for the chosen interatomic interaction
in \cite{Saeidian}. Specifically, in \cite{Saeidian} the effective range $R$ was not
included in the definition of the scattering volume $V_p$ and the
interatomic interaction was of single channel character, contrary to the
tensorial structure of $\hat{V}$ in (\ref{Tensor potential}).
We note that we have shown recently \cite{Saeidian2012} using a tensorial potential of type (\ref{Tensor potential})
that both these effects do not influence the position of the s-wave CIR.

\subsection{Shift of p-wave Feshbach resonance in harmonic
waveguide} 
In \cite{Granger} the CIR of a pair of identical fermions has been
studied analytically by means of a fermion-boson mapping. In this
way the authors have obtained the ratio
\begin{equation}\label{resonance position old formula}
\frac{V^{CIR}_p}{V_{\perp}}=[12\zeta(-\frac{1}{2},1-\epsilon)]^{-1}
\end{equation}
for the position $V^{CIR}_p$ of the CIR
corresponding to the minimum in the transmission $T$. This ratio
depends only on the rescaled longitudinal energy $\epsilon$ and the dependence is defined by the Hurwitz
zeta-function $\zeta$. In Fig.\ref{fig4} we show the
resonance position predicted by this formula as a function of
$\epsilon$ (the solid curve). On the other hand, in \cite{Kim3},
the dependence of the p-wave CIR position on $\epsilon$ is given
by the simple expression
\begin{equation}\label{resonance position Kim}
\frac{V^{CIR}_p}{V_{\perp}}=-[4(1-\epsilon)]^{-3/2}\,\,,
\end{equation}
predicting a smoother dependence on $\epsilon$ and a strong
deviation from (\ref{resonance position old formula})
with varying $\epsilon$ (see the dashed curve in
Fig.\ref{fig4}). Thus, both formulae exhibit a strong
dependence of the p-wave CIR position on $\epsilon$ in support of
our above result (see Fig.\ref{fig2}) but do not describe
the strong dependence of the ratio $V_p^{CIR}/V_{\perp}$
on the trap frequency $\omega_{\perp}$ we find: see
Fig.\ref{fig3} and Figs.10(a,b) in our previous work
\cite{Saeidian}. Our above analysis has shown that the origin of the
drawback of both formulae (\ref{resonance position old formula})
and (\ref{resonance position Kim}) is the neglecting of the second
term $\frac{k^2}{R(B)}$ in Eq.(\ref{scattering volume}) in the
derivation of these formulae.

In a recent work \cite{Peng2014} it was concluded that,
unlike in the case of s-wave interaction, the p-wave effective range is
essential in the strongly interacting regime i.e. near the corresponding Feshbach
resonance and the shift of the p-wave magnetic Feshbach
resonance in confining harmonic waveguide was calculated
analytically. Assuming that the external magnetic field lies in
the $xz$-plane with angle $\phi$ with respect to the $z$-axis, and
that the incoming particles are in the ground state of an axially
symmetric harmonic confinement, the authors of
\cite{Peng2014} have obtained an analytical expression for the
scattering wave function. They arrive at the energy
dependent 1D scattering length:
\begin{equation}\label{1d scattering lenght}
a_{1D}=-\frac{\tan\delta_1}{k_0}=\frac{6a_{\perp}(D^{1D}_{1x}\cos^2\phi+D^{1D}_{0x}\sin^2\phi)}{\bar{D}^{1D}_{0z}D^{1D}_{1x}\cos^2\phi+D^{1D}_{0x}\bar{D}^{1D}_{1z}\sin^2\phi}
\end{equation}
where $k_0=\sqrt{2\mu(E-\hbar\omega_{\perp})}/\hbar =
\frac{2\sqrt{\epsilon}}{a_{\perp}}$
($\epsilon=\frac{E}{2\hbar\omega_{\perp}}-\frac{1}{2}$),
$\bar{D}^{1D}_{mz}=D^{1D}_m-12\zeta(-\frac{1}{2},1-\epsilon)$,
$D^{1D}_{mx}=D^{1D}_m+C^{1D}_x(\epsilon)$,
$D^{1D}_m=\frac{V_{\perp}}{V_{p}}-\frac{2E}{\hbar\omega_{\perp}}\frac{a_{\perp}}{R}$
and $C^{1D}_x(\epsilon)$ is a function defined in \cite{Peng2014}.

For anisotropic interaction ($D^{1D}_0\neq D^{1D}_1$) with
$\phi=0$ or $\phi=\frac{\pi}{2}$ as well as for isotropic
interaction ($D^{1D}_0= D^{1D}_1$), one obtains:
\begin{equation}\label{1Dscattering length}
a_{1D}=\frac{6a_{\perp}}{\bar{D}^{1D}_{mz}}
\end{equation}
which yields a resonance ($\bar{D}^{1D}_{mz}=0$) at
$D^{1D}_m-12\zeta(-\frac{1}{2},1-\epsilon)=0$, from which one
obtains for the resonance condition i.e. for the position of the p-wave
CIR:
\begin{equation}\label{resonance position new formula}
\frac{V^{CIR}_p(\epsilon)}{V_{\perp}}=[\frac{2a_{\perp}}{R}+12\zeta(-\frac{1}{2},1-\epsilon)]^{-1}\,\,.
\end{equation}
This formula is essentially different from Eqs.(\ref{resonance
position old formula}) and (\ref{resonance position Kim}) by the
presence of the term
$\frac{2a_{\perp}}{R}=\frac{2\sqrt{\hbar}}{R\sqrt{\mu\omega_{\perp}}}$.

In our case of two spin-polarized $^{40}K$ atoms $|F=9/2,
m_F=-7/2\rangle$ with relative quantum numbers $l=1$ and $m_l=0$ trapped
in the waveguide and equipped with a longitudinal magnetic field
we expect the resonance position to obey Eq.(\ref{resonance
position new formula}). The results presented in Fig.\ref{fig5}
demonstrate excellent agreement of our calculation of the p-wave
CIR position $V_p^{CIR}/V_{\perp}$ with the analytic
result (\ref{resonance position new formula}) in the complete region
below the first threshold of transverse excitation $\epsilon=1$,
which persists in a broad range of varying $\omega_{\perp}$.
Note also the emerging stronger than linear dependence on the
energy $\epsilon$ with increasing $\omega_{\perp}$.

Fig.\ref{fig6} shows the resonance position as a function of
the trap frequency $\omega_{\perp}$ for the
zero energy limit $\epsilon=5\times 10^{-4}$ as well as near the first threshold
of transverse excitation $\epsilon=0.9$.  It demonstrates the close to linear dependence of the
resonance position $V_p^{CIR}/V_{\perp}$ on $\sqrt{\omega_{\perp}}$ as
$\omega_{\perp}\rightarrow 0$ regardless of the value of the energy (see figure inside).  The numerical results are in good
agreement with Eq.(\ref{resonance position new formula}).

\subsection{Width of p-wave Feshbach resonance}
Next we analyze the widths of the p-wave CIRs
in harmonic waveguides. Since at the resonant field $B_{CIR}$ (corresponding to the
position of CIR) the transmission reaches $T(B_{CIR})=0$ and the
maximal $T$ value is $1$ we define the width $\Delta_{1D}=B_+ -
B_-$, where $B_+$ and $B_-$ are the fields correspondingly right
and left to $B_{CIR}$ and the transmission approaches the value
$T(B_{\pm})=1/2$. This definition differs from that for a free space
magnetic Feshbach resonance where it reads $B^*-B_0$ \cite{Lange}. By using the analytical formula for the scattering
amplitude in ref. \cite{Shi2014}
\begin{equation}\label{scattering amplitude formula}
f_p(B)=\frac{-ik_0}{a_{1D}^{-1}(B)+Rk_0^2+iK_0}
\end{equation}
valid near the CIR in the zero-energy limit $k_0=\sqrt{2\mu
E_{\|}}/\hbar = \sqrt{2\mu(E-\hbar\omega_{\perp})}/\hbar
\rightarrow 0$ and Eq.(\ref{1Dscattering length}) we obtain
\begin{equation}\label{width formula}
\Delta_{1D}=\Delta[\frac{\alpha_+}{\alpha_+-1}-\frac{\alpha_-}{\alpha_--1}]
\end{equation}
where $\alpha_{\pm}=6\frac{V_{bg}}{a_{\perp}^2a_{1D}(B_{\pm})} +
\frac{V_{bg}}{V_p^{CIR}}$. In Figs. \ref{fig7} and \ref{fig8} we
show the calculated resonance width $\Delta_{1D}$ as a function of
the rescaled energy $\epsilon$ and waveguide frequency
$\omega_{\perp}$, respectively, together with the analytical result
according to Eq.(\ref{width formula}). One observes a good agreement
for $\omega_{\perp}< 0.01$. However, we encounter major deviations with
increasing $\omega_{\perp}$ except for the zero energy limit
$\epsilon\rightarrow 0$.  These deviations are due to the fact
that the analytical formula (\ref{scattering amplitude formula})
which we used to derive (\ref{width formula}), was obtained in the zero energy limit and
does not work for larger energies. In Fig.\ref{fig9} we present the real part of our scattering amplitude $Re(f_p)$ along with the analytical results from (\ref{scattering amplitude formula}) as a function of $V_p/V_{\perp}$ for $\epsilon=0.9$ (large energy) for comparison.  As the trap frequency $\omega_{\perp}$ increases we observe
deviations between the analytical and numerical results for $f_p$.

The same way as in the case of the free space resonance, the width
$\Delta_{1D}$ of the CIR narrows with decreasing energy
$\epsilon$ (see Fig.\ref{fig7}). Fig.\ref{fig8} shows that in a
harmonic waveguide there is a region where we have a possibility
for narrowing the width by decreasing the trap frequency.

\section{Conclusion}
We develop and analyze a theoretical model to study Feshbach
resonances of identical fermions in atomic waveguides by extending
the two-channel model suggested in \cite{Lange}  and adopted in
\cite{Saeidian2012} for confined bosons. In this model, the
experimentally known parameters of Feshbach resonances in free
space are used as an input. Within this approach we have
calculated the shifts and widths of p-wave magnetic Feshbach
resonance of $^{40}$K atoms in the hyperfine state $|F=9/2,
m_F=-7/2\rangle$ and for the relative angular momentum state
$|l=1, m_l=0\rangle$ emerging in harmonic waveguides as p-wave
CIRs.  We find a linear dependence of the resonance position on
the longitudinal colliding energy below the threshold for the
first transverse excitation. It is shown that in a harmonic
waveguide there is the possibility to decrease the width of the
p-wave Feshbach resonance by decreasing the (transversal) trap
frequency which could be used in corresponding experiments. Our
analysis demonstrates the importance of including the effective
range terms in the computational schemes for the description of
the p-wave CIRs contrary to the case of s-wave CIRs where the
impact of the effective radius is negligible. In previous
investigations of the p-wave CIRs in harmonic waveguides
\cite{Granger,Kim,Saeidian} the effects due to the effective range
have been neglected. The developed model can be applied for a
quantitative analysis of other p-wave CIRs following a different
spin structure and for confining traps of different geometry
including effects due to anharmonicity and anisotropy.

\section{ACKNOWLEDGEMENTS}
Sh.S would like to thank J. Abouie and S. Abedinpour for
fruitful discussions. V.S.M.and P.S. acknowledge financial support by the
Heisenberg-Landau Program.
V.S.M. thanks the Zentrum f\"ur Optische Quantentechnologien of
the University of Hamburg and Sh.S. thanks the Bogoliubov
Laboratory of Theoretical Physics of JINR for their warm
hospitality. This work was supported by IASBS (Grant No.G2014IASBS12648).

\end{document}